\documentclass[12pt]{article}
\usepackage{graphicx, url}
\usepackage{palatino}
\usepackage{amsmath, amssymb,bbm}
\usepackage{epsfig}
\usepackage{rotating}
\usepackage{dcolumn}
\usepackage{bm}
\usepackage{caption}
\usepackage{subcaption}
\usepackage[table]{xcolor}
\newcommand{\comment}[1]{}
\def \non{\nonumber}
\def \ra{\rightarrow}
\def \bea{\begin{eqnarray}}
\def \eea{\end{eqnarray}}

\def \ccbar{c\overline{c}}

\def \ppbar{p\overline{p}}

\def \sg{\sigma}
\begin{document}   
\baselineskip 18pt
\title{Probing charmonia resonances in proton-lead collisions at the LHC}
\author{
   Sudhansu~S.~Biswal$^1$\footnote{E-mail: ssbiswal@ravenshawuniversity.ac.in}, 
   ~Sushree~S.~Mishra$^1$\footnote{E-mail: sushreesimran.mishra97@gmail.com},\\ 
   ~Monalisa Mohanty$^1$\footnote{E-mail: monalimohanty97@gmail.com}
     ~and  K.~Sridhar$^2$\footnote{E-mail: sridhar.k@apu.edu.in} \\ [0.2cm]
    {\it \small 1. Department of Physics, Ravenshaw University,} \\ [-0.2cm]
    {\it \small Cuttack, 753003, India.}\\ [-0.2cm]
    {\it \small 2. School of Arts and Sciences, Azim Premji University,} \\ [-0.2cm]
    {\it \small Sarjapura, Bangalore, 562125, India.}\\
}
\date{}
\maketitle

\begin{abstract}

After achieving a successful description of charmonia production in proton-proton 
(pp) collisions within Non-Relativistic Quantum Chromodynamics (NRQCD) and 
modified NRQCD (MNRQCD) frameworks, we now extend our study to proton-lead (p-Pb) 
collisions at the Large Hadron Collider (LHC). 
In this context, theoretical predictions for the 
production cross-section, the nuclear modification factor $R_{pPb}$ and 
forward -to-backward production 
ratio $R_{FB}$ of $J/\psi$ have been made as a function of 
transverse momentum and rapidity at the LHC. MNRQCD provides a reasonable 
agreement with the available LHC data for $J/\psi$ production.
Furthermore, using the 
heavy quark symmetry relation, we have estimated the production 
cross sections and expected number of 
events of $\eta_{c}$ and $h_{c}$ production in both NRQCD and MNRQCD. 
Notably, the predictions exhibit substantial differences in the integrated 
cross-sections as well as in the $p_T$ distributions for these states at the LHC, 
highlighting the importance of further experimental and theoretical studies of 
these resonances to advance our understanding of quarkonium production.

\end{abstract}

\maketitle

\noindent
Three decades ago, it was proposed that the formation of Quark - Gluon Plasma 
(QGP) \cite{Matsui:1986dk, Pasechnik:2016wkt, Blaizot:2015hya} would reduce the 
yield of $J/\psi$ mesons in high-energy heavy ion collisions compared 
to proton-proton (pp) collisions. 
This suppression was attributed to the Debye screening of the 
heavy-quark potential at finite temperatures. 
This predicted signature of QGP shows 
significant experimental and theoretical research 
into the production of heavy quarkonium in nuclear collisions.
The Relativistic Heavy Ion Collider (RHIC) \cite{PHENIX:2004vcz,STAR:2005gfr} 
and the Large Hadron Collider (LHC) 
are the best experimental facilities that 
are designed to investigate an exotic state of matter
such as QGP which has been explored at both perturbative
and non-perturbative sector of QCD.
The investigation of heavy quarkonium 
bound states in heavy-ion collisions \cite{Lan, Brambilla} 
play a crucial role in exploring the properties of the deconfined QGP. 
This study examines proton-lead (p-Pb) collisions, 
where the behavior of heavy quarkonium bound states offer valuable insights 
into the formation and properties of the QGP.

In contrast to pp collisions, quarkonium production in p-Pb collisions is affected 
of cold nuclear matter effects, primarily arising from modifications of the 
parton distributions inside the nucleus. These lead to deviations in the 
production cross-sections, quantified through the nuclear modification factor 
$R_{pPb}$ and result in a forward-to-backward ratio, 
making p-Pb collisions a crucial probe to 
distinguish nuclear effects from the pp production.

In relativistic heavy-ion collisions, charm quark pairs are 
generated during the early stages of the interaction, 
but their conversion into charmonium bound states occur more gradually. 
This distinction underscores the complex 
dynamics of particle production and hadronization in 
heavy-ion collisions. In particular, 
$J/\psi$ meson production serves as a crucial observable for 
studying the QGP \cite{kha}
- a high-temperature, deconfined state of matter where quarks and gluons exist freely, 
rather than being confined within hadrons. 
Some of the theoretical predictions in heavy ion 
collisions \cite{Gavai, Satz, Basu, Zhou, Du, Andronic} have also been tested. 
To improve our understanding of the QGP, some analysis in 
proton-proton (pp) and proton-nucleus (p-A) collisions are necessary.

Charmonia serve as particularly powerful probes for studying the QGP.
The production of charmonia can indeed be effectively studied using 
Non-Relativistic Quantum Chromodynamics (NRQCD) \cite{bbl}, 
which is a powerful theoretical framework that simplifies the calculations of heavy 
quarkonium production and dynamics by exploiting the non-relativistic nature of the 
heavy quark and anti-quark when their relative velocity 
is small compared to the speed of light.

In NRQCD at leading order, the  $Q\bar Q$ state is in a 
color-singlet state, however at $O(v)$, 
it can be in a color-octet state which is connected to the physical 
quarkonium state through a non-perturbative gluon emission.
The cross-section for the production of a 
quarkonium state $H$ of mass $M$ in NRQCD can be expressed as:

\bea
  \sigma(H)\;=\;\sum_{n=\{\alpha,S,L,J\}} {F_n\over {M}^{d_n-4}}
       \langle{\cal O}^H_n({}^{2S+1}L_J)\rangle, 
\label{factorizn}
\eea
where $F_n$'s are the short-distance coefficients 
and ${\cal O}_n$ are operators of naive dimension $d_n$, 
describing the long-distance effects. Due to NRQCD factorization, 
the non-perturbative matrix elements are energy independent 
and can be extracted at a given energy and used in the 
prediction of quarkonium cross-sections at other energies.

NRQCD has been more successful in explaining the systematics 
of quarkonium production at the Fermilab Tevatron~\cite{cdf,CDF:2001fdy}, 
compared to the then existing 
Color Singlet Model (CSM) \cite{br}, which was used to analyze 
the production of quarkonia, where the $Q \bar Q$ state produced 
in the short-distance process was assumed to be a color-singlet.
NRQCD predicts transverse polarisation for quarkonia production at high $p_T$, 
but experiments fail to see any evidence for the polarisation 
in $J/\psi$~\cite{jpsi_pol} or $\Upsilon$~\cite{CDF:2001fdy, CMS1, LHCb2} measurements.
Therefore, independent tests of 
NRQCD~\cite{Sridhar:1996vd,Sridhar:2008sc,tests,Mathews:1998nk,bs1,test1} 
are consequently important and the prediction of polarisation of the produced
quarkonium state is an important test.

In Refs.~\cite{bms12, bms3, bms4, bms5}, 
we have studied quarkonia production, where we get a 
significant difference between the NRQCD and 
modified NRQCD (MNRQCD) model in case of $\eta_c$, $\eta_b$, 
$h_c$ and $h_b$ production. 
However, MNRQCD predictions for $\eta_c$ and $h_c$ production 
shows good agreement with LHCb experimental results. 
Motivated by the success of MNRQCD in 
the proton-proton collisions, we have extended our 
work to study charmonia production in heavy ion collisions. 
In this paper, we have focused our analysis on charmonia production in 
proton-lead collisions using both NRQCD and MNRQCD models.

The NRQCD formula for $J/\psi$ can be written explicitly in terms 
of the various octet and singlet intermediate states:
\begin{eqnarray}
\sigma_{J/\psi}  = \hat F_{{}^{3}S_1^{[1]}} \times \langle {\cal O} ({}^{3}S_1^{[1]}) \rangle +
                \hat F_{{}^{3}S_1^{[8]}} \times \langle {\cal O} ({}^{3}S_1^{[8]}) \rangle +\cr
                 \hat F_{{}^{1}S_0^{[8]}} \times \langle {\cal O} ({}^{1}S_0^{[8]}) \rangle 
                + {1 \over M^2} \biggl\lbrack\hat F_{{}^{3}P_J^{[8]}} \times \langle {\cal O} 
                     ({}^{3}P_J^{[8]}) \rangle \biggr\rbrack .
\label{Fock}
\end{eqnarray}

The above formula gets modified to the following in the MNRQCD with perturbative soft gluon emission:

\begin{eqnarray}
\sigma_{J/\psi} &=& \biggl\lbrack \hat F_{{}^{3}S_1^{[1]}} 
                \times \langle {\cal O} ({}^{3}S_1^{[1]}) \rangle \biggr\rbrack \cr 
                &+& \biggl\lbrack  
                  \hat F_{{}^{3}S_1^{[8]}} 
                 + \hat F_{{}^{1}P_1^{[8]}} 
                + \hat F_{{}^{1}S_0^{[8]}} + (\hat F_{{}^{3}P_J^{[8]}} ) \biggr\rbrack 
                \times ({\langle {\cal O} ({}^{3}S_1^{[1]}) \rangle \over 8}) \cr
                &+& \biggl\lbrack  
                  \hat F_{{}^{3}S_1^{[8]}} 
                 + \hat F_{{}^{1}P_1^{[8]}} 
                + \hat F_{{}^{1}S_0^{[8]}} + (\hat F_{{}^{3}P_J^{[8]}} ) \biggr\rbrack 
                \times \langle {\cal O}  \rangle ,
\end{eqnarray}
where
\begin{equation}
     \langle {\cal O}  \rangle =
                \times \biggl\lbrack 
                 \langle {\cal O} ({}^{3}S_1^{[8]}) \rangle 
                + \langle {\cal O} ({}^{1}S_0^{[8]}) \rangle 
                + {\langle {\cal O} ({}^{3}P_J^{[8]}) \rangle \over M^2}
                    \biggr\rbrack. 
\end{equation}

The differential cross section for $c\bar c$ pair production with specific 
angular momentum and color states at the LHC is given by:

\bea
&&\frac{d\sg}{dp_{_T}} \;(p  ~Pb \ra \ccbar\; [^{2S+1}L^{[1,8]}_J]\, X)= \non \\
&&\sum \int \!dy \int \! dx_1 ~x_1\:G_{a/p} (x_1)~x_2\:G_{b/Pb}(x_2) 
\:\frac{4p_{_T}}{2x_1-\overline{x}_T\:e^y}\non\\
&&\frac{d\hat{\sg}}{d\hat{t}}
(ab\ra \ccbar[^{2S+1}L_J^{[1,8]}]\;d),
\label{eq:diff}
\eea
where the summation is over the partons ($a$ and $b$),    
the final state $c\bar c$ is in the $^1 S^{[8]}_0$, $^1 P^{[8]}_1$,  
$^3 S^{[8]}_1$ states and $G_{a/p}$,   
$G_{b/Pb}$ are the distributions of partons $a$ and 
$b$ in the proton and lead respectively. Here, $x_1$ and $x_2$ are the respective  momentum they carry. 
In the above formula, 
$\overline{x}_T=\sqrt{x_T^2+4\tau} \equiv 2 M_T/\sqrt{s}$ \ with 
\  $x_T=2p_{_T}/\sqrt{s}$ and  \(\tau=M^2/s\).
$\sqrt{s}$ is the center-of-mass energy, $M$ is the mass of the resonance 
and $y$ is the rapidity at which the resonance is produced. 
Here parton distributions for the nucleus are as follows \cite{Kovarik}:
\bea
 G_{b/Pb}(x_{2}) = \dfrac{Z}{A}G^{p/A}(x_{2})+\dfrac{A-Z}{A}G^{n/A}(x_{2}).
\eea
The fixed-order perturbative calculation
have been used to get the cross-section for charmonia production 
and a cut-off is imposed in the calculations for
low-$p_T$ regime. The charmonia cross-section in the low-$p_T$ 
region requires a resummation of multiple gluon radiation.

\begin{figure}[!h]
\begin{center}
\includegraphics[width=14cm, height=7.5cm]{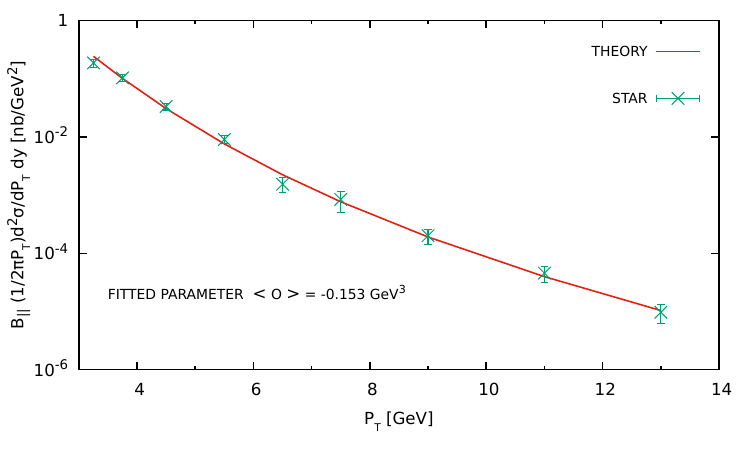}\caption{Theoretical prediction of 
differential cross sections fitted to the data on $J/\psi$ production from the STAR experiment at RHIC.}
        \label{fig:fig1}
\end{center}
\end{figure}

 Fig. \ref{fig:fig1} shows the fit to the 200 GeV RHIC \cite{STAR} data on $J/\psi$ production, 
where $B_{||}$ is the branching ratio for $J/\psi \rightarrow e^+ e^-$
(5.971\%), including branching ratio for $J/\psi \rightarrow \gamma e^+ e^-$ (0.88\%).

\begin{figure}[!h]
\begin{center}
\includegraphics[width=15cm, height=9cm]{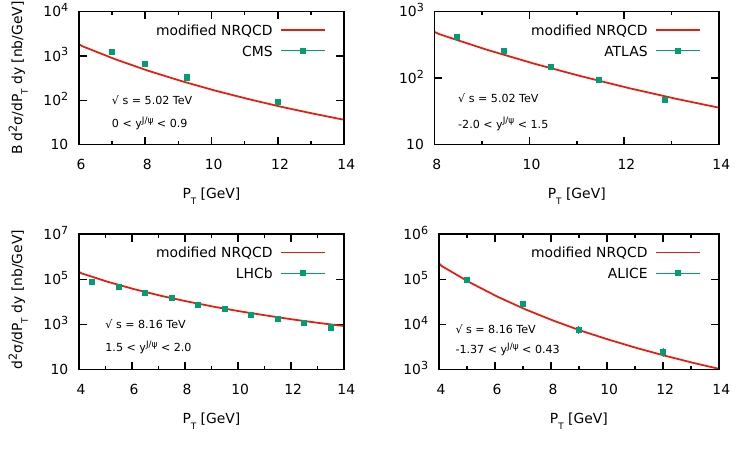}
\caption{Predicted differential distributions for $J/\psi$ 
production compared with the data from the LHC experiments.}
        \label{fig:fig2}
\end{center}
\end{figure}

\begin{figure}[!h]
\begin{center}
\includegraphics[width=15.5cm, height=8cm]{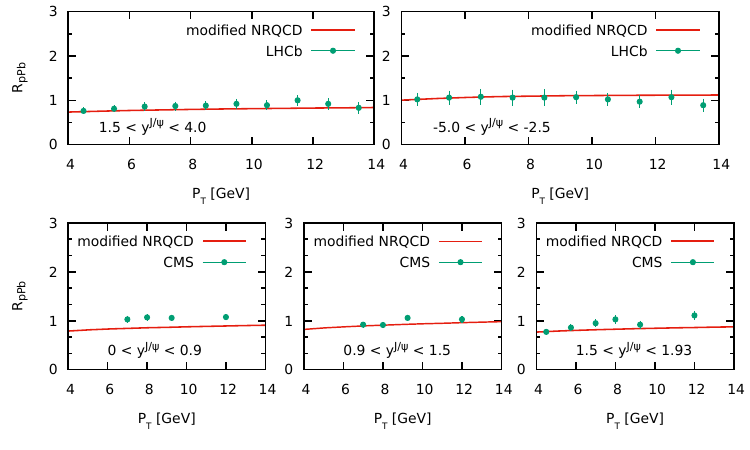}
\caption{Transverse momentum dependence of $R_{pPb}$ for 
$J/\psi$ production at the LHC.}
        \label{fig:fig3}
\end{center}
\end{figure}

\begin{figure}[!h]
\begin{center}
\includegraphics[width=15.5cm, height=8.5cm]{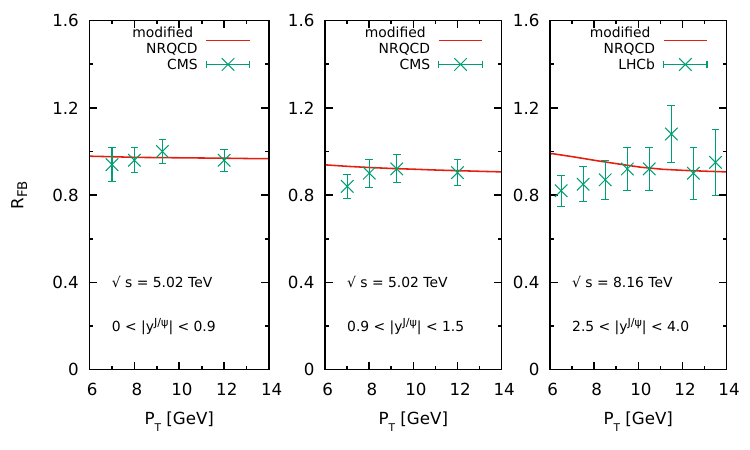}\caption{Transverse momentum 
dependence of $R_{FB}$ for $J/\psi$ production at the LHC.}
        \label{fig:fig4}
\end{center}
\end{figure}

In Fig. \ref{fig:fig2}, the theoretical production cross-sections for $J/\psi$ 
have been compared with the LHC \cite{CMS,LHCbf, ALICEf, ATLASf} 
experimental data, showing an agreement with MNRQCD for proton-lead collisions. 
Here while comparing with the CMS and ATLAS experimental results, we have taken the 
branching ratio for $J/\psi \rightarrow \mu^+ \mu^-$ for 
theoretical predictions of $J/\psi$ production.

Nuclear modification factor is defined as the ratio of the yield in heavy-ion collisions 
to that in proton-proton collisions, scaled by the average number of binary collisions. 
Here the nuclear modification factor for pPb configuration is:

\begin{equation}
      R_{pPb}(p_T, |y|) = \frac{1}{208} \frac{d^2\sigma_{pPb}/dp_{T}dy}{d^2\sigma_{pp}/dp_{T}dy},
\end{equation}

where, $\sigma_{pp}$ is the cross-section from pp collisions.

The formula for forward-to-backward ratio is defined as:

\begin{equation}
      R_{FB}(p_T, |y|) = \frac{d^2\sigma_{pPb}(p_T, +|y|)/dp_{T}dy}{d^2\sigma_{pPb}(p_T, -|y|)/dp_{T}dy}.
\end{equation}

Fig. \ref{fig:fig3} and Fig. \ref{fig:fig4} show the nuclear modification factor 
and forward-to-backward production ratio predictions using MNRQCD model, 
which gives significant results while comparing with the experimental data. 
The fig. \ref{fig:fig3} and fig. \ref{fig:fig4} suggest
that $R_{pPb}$ and $R_{FB}$ work well as an observable up to at least 
$p_T \sim$ 14~GeV, exhibiting a stable behavior around unity.

\begin{figure}[h]
\begin{center}
\includegraphics[width=12.5cm, height=7.2cm]{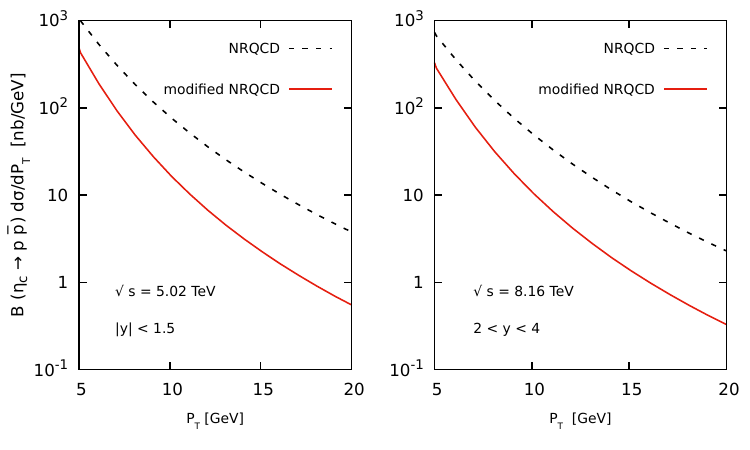}
\caption{ Predicted differential cross-sections for $\eta_{c}$ production at the LHC. }
        \label{fig:fig5}
\end{center}
\end{figure}

\begin{figure}[!h]
\begin{center}
\includegraphics[width=12.5cm, height=7.2cm]{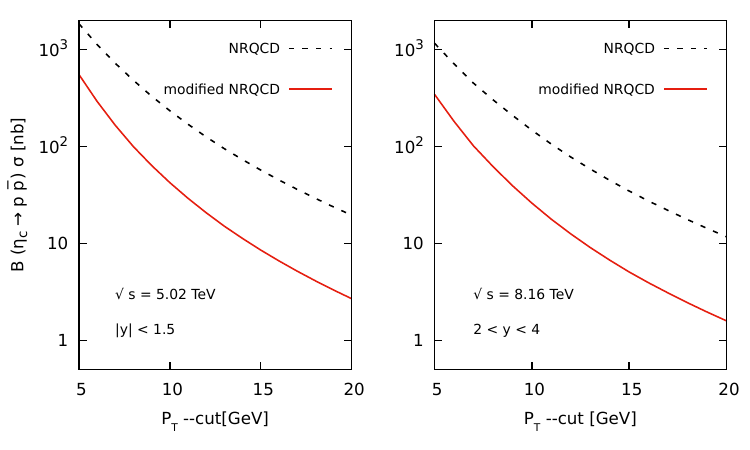}
\caption{ Predicted Integrated cross-sections for $\eta_{c}$ production at the LHC. }
        \label{fig:fig6}
\end{center}
\end{figure}

\begin{table}[htbp]
\centering
 \hspace*{-0.1cm}
\begin{tabular}{ |m{0.5cm}|m{0.5cm}|m{0.5cm}|m{0.5cm}|m{0.5cm}| }
 \hline
\multicolumn{1}{|c|}{ } &\multicolumn{4}{c|}{$\sim$Expected number of events} \\[1mm]

 \cline{2-5} 
\multicolumn{1}{|c|}{} &\multicolumn{2}{c|}{}
&\multicolumn{2}{c|}{} \\[-2mm]

\multicolumn{1}{|c|}{} &\multicolumn{2}{c|}{Rapidity: $|y| < 1.5$, $\sqrt{s}$ = 5.02 TeV}
&\multicolumn{2}{c|}{Rapidity: $2 < y < 4$, $\sqrt{s}$ = 8.16 TeV} \\[1mm]
\cline{2-5}
&&&& \\[-3mm]
\multicolumn{1}{|c|}{Model} 
	&\multicolumn{1}{c|}{$P_{T}$ $>$ 5 GeV}&\multicolumn{1}{c|}{$P_{T}$ $>$ 10 GeV}
&\multicolumn{1}{c|}{$P_{T}$ $>$ 5 GeV}&\multicolumn{1}{c|}{$P_{T}$ $>$ 10 GeV}
 \\[1mm]

\hline
\hline

\multicolumn{1}{|c|}{NRQCD} &
\multicolumn{1}{c|}{$1.8 \times 10^{5}$} &\multicolumn{1}{c|}{$2.3 \times 10^{4}$}
&\multicolumn{ 1}{c|}{$1.2 \times 10^{5}$}&\multicolumn{1}{c|}{$1.5 \times 10^{4}$}  
 \\
\hline
\multicolumn{1}{|c|}{MNRQCD} &
\multicolumn{1}{c|}{$5.5 \times 10^{4}$} &\multicolumn{1}{c|}{$4.2 \times 10^{3}$}
&\multicolumn{1}{c|}{$3.4 \times 10^{4}$} &\multicolumn{1}{c|}{$2.6 \times 10^{3}$}   
\\
 \hline 
  
\end{tabular}
\caption{\label{tab:events1}
Number of $\eta_c$ events expected at the LHC correspond to 
integrated luminosity of 100 nb$^{-1}$. 
}
\end{table}

\begin{figure}[!h]
\begin{center}
\includegraphics[width=12.5cm, height=7.0cm]{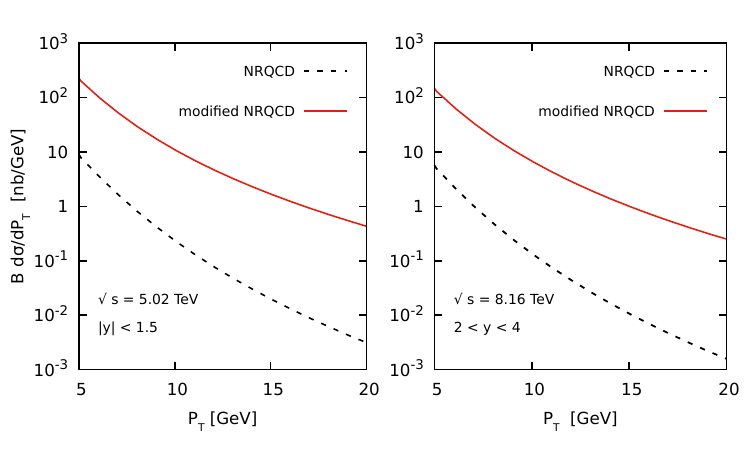}
\caption{ Predicted differential cross-sections for $h_{c}$ production at the LHC.}
        \label{fig:fig7}
\end{center}
\end{figure}

\begin{figure}[!h]
\begin{center}
\includegraphics[width=12.5cm, height=7.0cm]{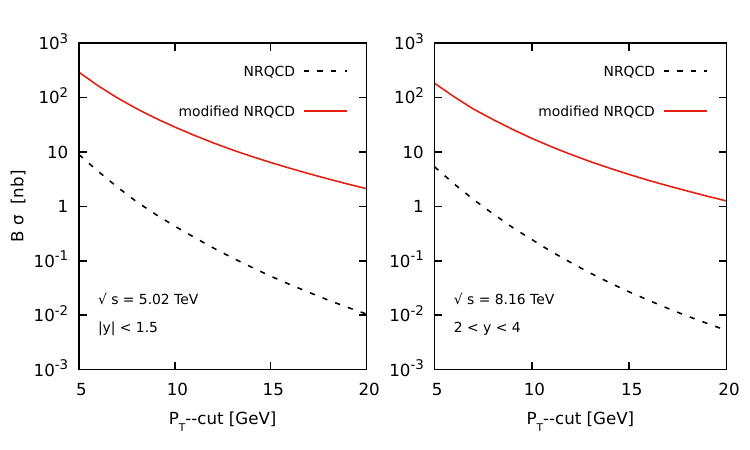}
\caption{ Predicted Integrated cross-sections for $h_{c}$ production at the LHC.}
        \label{fig:fig8}
\end{center}
\end{figure}

\begin{table}[htbp]
\centering
 \hspace*{-0.1cm}
\begin{tabular}{ |p{1cm}|p{1cm}|p{1cm}|p{1cm}|p{1cm}| }
 \hline
\multicolumn{1}{|c|}{ } &\multicolumn{4}{c|}{$\sim$Expected number of events} \\[2mm]

 \cline{2-5} 
\multicolumn{1}{|c|}{} &\multicolumn{2}{c|}{}
&\multicolumn{2}{c|}{} \\[-2mm]

\multicolumn{1}{|c|}{} &\multicolumn{2}{c|}{Rapidity: $|y| < 1.5$, $\sqrt{s}$ = 5.02 TeV}
&\multicolumn{2}{c|}{Rapidity: $2 < y < 4$, $\sqrt{s}$ = 8.16 TeV} \\[2mm]
\cline{2-5}
&&&& \\[-3mm]
\multicolumn{1}{|c|}{Model} 
	&\multicolumn{1}{c|}{$P_{T}$ $>$ 5 GeV}&\multicolumn{1}{c|}{$P_{T}$ $>$ 10 GeV}
&\multicolumn{1}{c|}{$P_{T}$ $>$ 5 GeV}&\multicolumn{1}{c|}{$P_{T}$ $>$ 10 GeV}
 \\[1mm]

\hline
\hline

\multicolumn{1}{|c|}{NRQCD} &
\multicolumn{1}{c|}{$9.3 \times 10^{2}$} &\multicolumn{1}{c|}{$45$}
&\multicolumn{ 1}{c|}{$5.5 \times 10^{2}$}&\multicolumn{1}{c|}{$25$}   
 \\

\hline

\multicolumn{1}{|c|}{MNRQCD} &
\multicolumn{1}{c|}{$3.0 \times 10^{4}$} &\multicolumn{1}{c|}{$3.0 \times 10^{3}$}
&\multicolumn{1}{c|}{$1.9 \times 10^{4}$} &\multicolumn{1}{c|}{$1.8 \times 10^{3}$}   
\\
 \hline   
\end{tabular}
\caption{\label{tab:events2}
Number of $h_c$ events expected at the LHC correspond to integrated luminosity of 100 nb$^{-1}$ . 
}
\end{table} 

For better understanding, we have extended our work on 
$p_T$ distributions of $\eta_c$ and $h_c$ production 
in both NRQCD and MNRQCD.
Figs. \ref{fig:fig5} and \ref{fig:fig6} represent the $\eta_c$ 
production differential cross-sections as a function
of $p_T$ and integrated cross-sections for different $p_T$ - cuts in both NRQCD and
MNRQCD. Similarly, Figs. \ref{fig:fig7} and \ref{fig:fig8} are for $h_c$ production.
We study $\eta_c$ in its decay 
into $\ppbar$ with a branching fraction of 1.33 $\times$ $10^{-3}$ \cite{ParticleDataGroup:2024cfk}. 
For $h_c$ production in its decay into $\eta_c$ and 
$\gamma$ with a branching fraction of 60$\%$ and $\eta_c$ in 
its decay into $\ppbar$ with a branching fraction 
of 1.33 $\times$ $10^{-3}$. 
As can be seen, both NRQCD and MNRQCD show significant deviations 
from their respective predictions for $\eta_c$ as well as $h_c$ production.

To get a sense of the feasibility of measuring the $\eta_c$ and $h_c$ production at the LHC, 
we have calculated the $p_T$ -integrated cross sections in two different rapidity ranges. 
These results, presented in tables 1 and 2 for an integrated luminosity of 100 nb$^{-1}$, 
suggest that a significant number of $\eta_c$ and $h_c$ events can be expected at the LHC.

In conclusion, we have studied $J/\psi$ production in MNRQCD for p-Pb collisions. 
We have fitted our model predictions to STAR data on $J/\psi$ production and used the 
fitted parameters to predict the distributions at the LHC energy, and find an 
agreement with the data from the CMS, LHCb, ALICE and ATLAS experiment. 
The nuclear modification factors and forward-to-backward ratios for $J/\psi$ show an agreement, 
when compared with the data from the LHC experiment.  
Furthermore, we have made predictions for $\eta_{c}$ and $h_c$ production using both NRQCD and 
MNRQCD, which give an insight to the charmonia production in heavy ion collisions.
In future, if the measurements of $\eta_{c}$ and $h_c$ become available, 
they will be experimentally challenging due to the difficulty in reconstructing 
its decay channels, which effectively leads to a suppressed production cross section.
However, with upcoming detector upgrades at the LHC and the HL-LHC, such measurements
may become feasible, offering a valuable opportunity 
to discriminate between NRQCD and MNRQCD predictions
and thereby deepen our understanding of heavy-quarkonium production.

\section*{Acknowledgments}
One of us (K.S.) gratefully acknowledges a research grant 
(No. 122500) from the
Azim Premji University. In this paper, as in everything else he writes, 
the results, opinions and views expressed are K.S.'s own and 
are not that of the Azim Premji University.


\end{document}